\documentclass[twocolumn,preprintnumbers,amssymb,amsmath,superscriptaddress,letterpaper,nofootinbib]{revtex4}[12pt]
\usepackage{times,graphics}
\usepackage[colorlinks]{hyperref}

\usepackage{graphicx}
\usepackage{dcolumn}
\usepackage{bm}
\usepackage{natbib}
\usepackage{graphicx}
\usepackage{epsfig}
\usepackage{epstopdf}
\usepackage{multirow}
\usepackage{amsmath}
\usepackage{lipsum}

\newcommand{\physrep}{{Physics~Reports}}

\newcommand{\apjs}{{Astrophys.~J.~Supp.}}

\newcommand{\mnras}{{Mon.~Not.~R.~Astron.~Soc.}}

\newcommand{\be}{\begin{equation}}
\newcommand{\ee}{\end{equation}}
\newcommand{\bea}{\begin{eqnarray}}
\newcommand{\eea}{\end{eqnarray}}

\makeatletter
\newcommand*{\shifttext}[2]{%
	\settowidth{\@tempdima}{#2}%
	\makebox[\@tempdima]{\hspace*{#1}#2}%
}
\makeatother

\begin{document}

\title{A Bayesian comparison between $\Lambda$CDM and phenomenologically emergent dark energy models }

\author{M. Rezaei}
\affiliation{Department of Physics, Bu-Ali Sina University, Hamedan
	65178, Iran}
\author{T. Naderi}
\affiliation{Department of Physics, Bu-Ali Sina University, Hamedan
	65178, Iran}
\author{M. Malekjani}
\email{malekjani@basu.ac.ir}
\affiliation{Department of Physics, Bu-Ali Sina University, Hamedan
	65178, Iran}
\author{A. Mehrabi}
\affiliation{Department of Physics, Bu-Ali Sina University, Hamedan
	65178, Iran}

\date{\today}

\begin{abstract}
In this work we examine the recently proposed phenomenological emergent dark energy (PEDE) model by \cite{Li:2019yem}, using the latest observational data in both expansion and perturbation levels. Applying the statistical Bayesian evidence as well as the AIC and BIC information criteria, we compare the PEDE model with the concordance $\Lambda$CDM model in both flat and non-flat universes. We combine the observational datasets as (i) expansion data (except CMB), (ii) expansion data (including CMB) and (iii) expansion data jointed to the growth rate dataset. Our statistical results show that the flat- $\Lambda$CDM model is still the best model. In the case of expansion data (including CMB), we observe that the flat- PEDE model is well consistent with observations as well as the concordance $\Lambda$CDM universe. While in the cases of (i) and (iii), the PEDE models in both of the flat and non-flat geometries are not favored. In particular, we see that in the perturbation level the PEDE model can not fit the observations as equally as standard $\Lambda$CDM cosmology. As the ability of the model, we show that the PEDE models can alleviate the tension of Hubble constant value appearing between the local observations and Planck inferred estimation in standard cosmology.

\end{abstract}
\maketitle

\section{Introduction}
Since the discovery of cosmic accelerated expansion, a flat Friedman-Roberson-Walker universe dominated by cold dark matter (CDM) and cosmological constant ($\Lambda$) has been introduced as the preferred model by cosmologists. This model is successful to explain many of cosmic observations including those of the Cosmic Microwave Background (CMB)\citep{Komatsu2011,Ade:2015rim}, Type-Ia Supernovae  (SnIa)\citep{Perlmutter1999,Kowalski2008}, Baryon Acoustic Oscillations (BAO) \citep{Eisenstein2005,Percival2010,Reid:2012sw} and $H(z)$ observations \citep{Sharov:2014voa}. However, the $\Lambda$CDM model suffers from some theoretical and observational problems. Theoretical problems include the fine-tuning (i.e., the fact that the value of this cosmological constant inferred from observations is extremely small 
compared with the energy scales	of high energy physics (Planck, grand unified theory, strong and even electroweak scales) and cosmic coincidence (why this kind of exotic matter starts to dominate today) issues \citep{Weinberg:1988cp,Carroll:2000fy, Padmanabhan2003, Copeland2006}. From the observational point of view, the $\Lambda$CDM cosmology plagued with some significant tensions in estimation of some key cosmological parameters. In particular, there is a statistically significant disagreement between the value of Hubble constant measured by the classical distance ladder and that of the Planck CMB data \citep{Freedman:2017yms}. Quantitatively speaking, we have $H_0=74.03\pm 1.42$ km/s/Mpc from the Cepheid-calibrated SnIa \citep{Riess:2019cxk}, while the $\Lambda$CDM cosmology deduced from Planck CMB data predicts $H_0 = 67.4\pm0.5$ km/s/Mpc \citep{Aghanim:2018eyx}. Also, the Lyman-$\alpha$ forest measurement of the Baryon Acoustic Oscillations obtained by BOSS in \citep{Delubac:2014aqe}, prefers a smaller value of the matter density parameter ($\Omega_{\rm m}$) compared to the value obtained by CMB data. Another tension concerns the discrepancy between large scale structure formation data \citep{Macaulay:2013swa}, and too large value of $\sigma_8$ predicted by the $\Lambda$CDM. The other observational problem regarding to $\Lambda$CDM model is the high tensions between cosmographic parameters of $\Lambda$CDM model and those of obtained from low-redshift observations \citep{Lusso:2019akb,Lin:2019htv}.

In order to overcome or at least alleviate the above problems, different kinds of dynamical dark energy (DE) models have been proposed. Many of review articles with comprehensive discussion on different aspects of various DE models are there in literature. Quintessence \citep{Erickson:2001bq}, ghost \citep{Veneziano1979,Rosenzweig:1979ay}, holographic \citep{Thomas2002},
k-essence \citep{Armendariz2001}, phantom \citep{Caldwell2002}, tachyon \citep{Padmanabhan2002}, dilaton \citep{Gasperini2002}, quintom\citep{Elizalde:2004mq} and dynamical vacuum energy \citep{Gomez-Valent:2014fda}  are examples of such dynamical DE models.
Moreover, many of these models have been compared with various observational data obtained from different cosmic surveys. In this procedure some of DE models have been ruled out and many of them achieve good consistency with observations \citep[see also][]{Malekjani:2016edh,Rezaei:2017yyj,Malekjani:2018qcz,Rezaei:2019roe,Rezaei:2019xwo}. Recently, a radical phenomenologically emergent DE model (PEDE) with symmetrical behavior around the current time has been proposed in \citep{Li:2019yem}. For this model at higher redshifts, DE has no effective role, while it emerges at later times. The interesting feature regarding PEDE is that, this model has no  degree of freedom, like the concordance $\Lambda$CDM model. In \citep{Li:2019yem}, authors by assuming hard cut priors from local measurement of the Hubble constant ,compared the model with combined sets of observations at both low and high
redshifts, include SnIa data, BAO data and CMB measurement. Their investigation showed that the PEDE model statistically is better than the $\Lambda$CDM cosmology. They concluded that the PEDE model can significantly reduce the tensions in estimation of the cosmological parameters, although some level of tension remains, in particular in estimation of the matter density. It can be useful to confront the PEDE model to other cosmological observations using different statistical methods. So, in this work we focus on the PEDE model and confront it with combination of different observational data sets using the Bayesian evidence method  as a most useful statistical analysis in modern cosmology. We will compare the PEDE model with the standard $\Lambda$CDM cosmology in the light of latest observational data. We organize the paper as follows. Firstly, we briefly introduce the PEDE model in Sec. (\ref{sec:de}). In Sec.(\ref{sec:be}), we present the cosmological data  as well as the Bayesian evidence analysis used in this work. In Sec. (\ref{sec:rd}), we present the main results of our work and provide the observational constraints on the model parameters. Finally, we conclude in Sec. (\ref{sec:c}).

\section{Phenomenological emergent dark energy versus $\Lambda$CDM}\label{sec:de}
Here, we introduce the PEDE model in standard cosmology and explain the difference between the behaviors of PEDE with concordance $\Lambda$CDM model in both background and cluster levels. In the context of standard gravity, adopting the Friedmann - Lemaître - Robertson - Walker (FLRW) metric, a general non-flat, isotropic and homogeneous universe can be explained by:

\begin{eqnarray}\label{eq:hub}
H^2 + \frac{K}{a^2}=\dfrac{8\pi G}{3}(\rho_{\rm r}+\rho_{\rm m}+\rho_{\rm d})&=& \nonumber\\\dfrac{8\pi G}{3}\left[ \rho_{\rm r0}(1+z)^4+\rho_{\rm m0}(1+z)^3+\rho_{\rm d0}f(z)\right]\;.
\end{eqnarray}
where subscript "0" indicates present values of parameters and $f(z)$ specifies the redshift evolution of $\rho_{\rm d}$. Using the dimensionless cosmological parameter $\Omega_{\rm i}=8\pi G \rho_{\rm i}/3H^2$, we can rewrite Eq.(\ref{eq:hub}) in the following form:

\begin{eqnarray}\label{eq:hub2}
\dfrac{H^2(z)}{H_0^2}= \Omega_{\rm r0}(1+z)^4 + \Omega_{\rm m0}(1+z)^3 - \Omega_{\rm K0}(1+z)^{2} +\nonumber \\ \Omega_{\rm d0}f(z)\;,
\end{eqnarray}

where $\Omega_{\rm K0}=K/H_0^2$ is the dimensionless curvature parameter. In the case of flat universe, Eq.(\ref{eq:hub2}) reduces to:

\begin{eqnarray}\label{eq:hub3}
\dfrac{H^2(z)}{H_0^2})= \Omega_{\rm r0}(1+z)^4 + \Omega_{\rm m0}(1+z)^3 + (1-\Omega_{\rm m0})f(z)\;.
\end{eqnarray}

Setting $f(z)=1$ in the Eqs.(\ref{eq:hub2}) \& (\ref{eq:hub3}), respectively, leads to non-flat and flat $\Lambda$CDM model. In PEDE cosmology, the density of DE reads \citep{Li:2019yem}:
\begin{eqnarray}\label{eq:omegd}
\Omega_{\rm d}(z)=\Omega_{\rm d0}\left[ 1-\tanh (\log (1+z))\right]\;.
\end{eqnarray}  
By using Eq.(\ref{eq:omegd}), the Hubble expansion in the case of PEDE model reads
\begin{eqnarray}\label{hub4}
\dfrac{H^2(z)}{H_0^2}=\nonumber \\ \sqrt{\Omega_{\rm r0}(1+z)^4+\Omega_{\rm m0}(1+z)^3-\Omega_{\rm K0}(1+z)^2+\Omega_{\rm d}(z)}\;.
\end{eqnarray}

\begin{figure} 
	\centering
	\includegraphics[width=8cm]{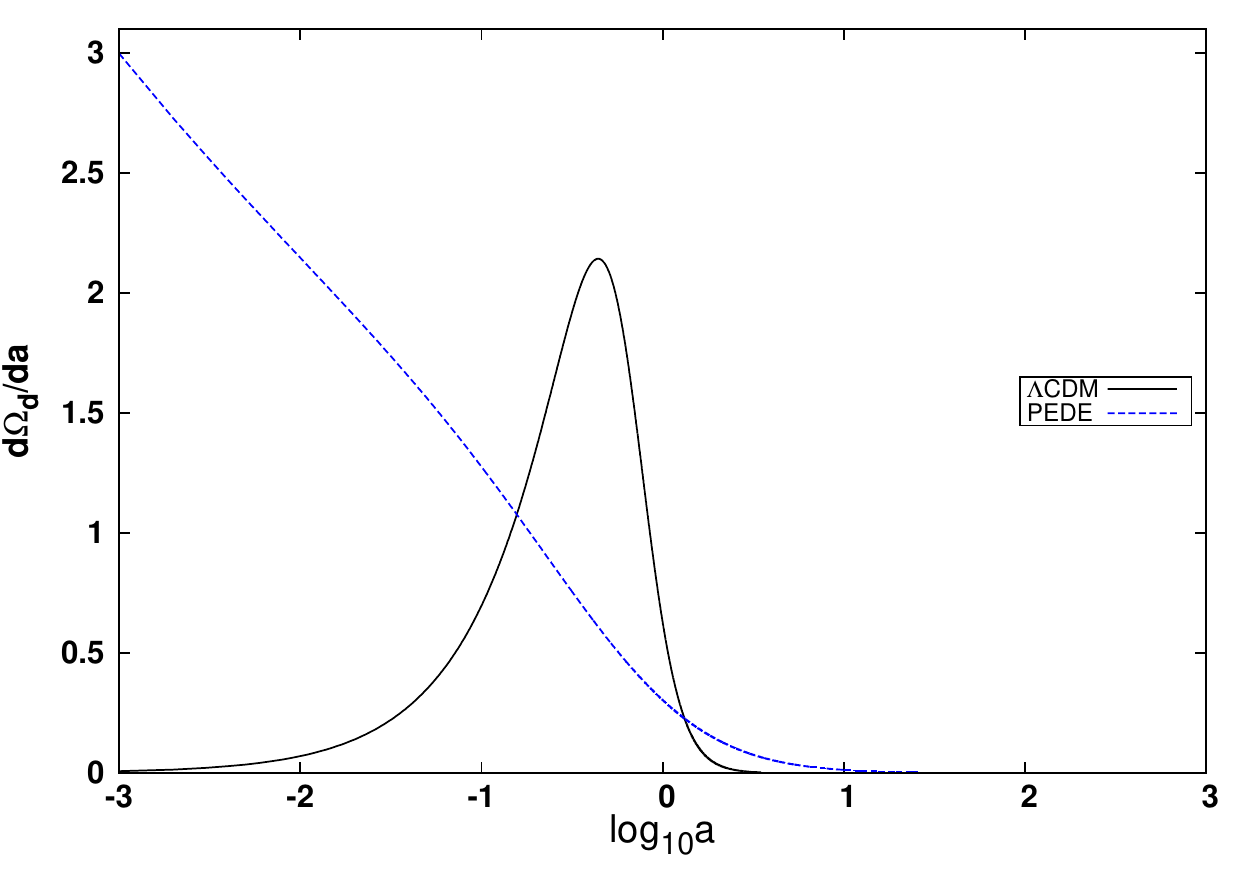}
	\caption{ The evolution of $ d\Omega_{\rm d}/da$, the rate of changes of  $\Omega_{\rm d}$, as a function of the scale factor $a$. We set $\Omega_{\rm m,0} = 0.3$.}
	\label{fig:figcoin}
\end{figure}
Now, we compare the PEDE model with $\Lambda$CDM one from the view point of coincidence problem. Based on the $\Lambda$CDM cosmology, at the early times, DE is negligible in comparison to other components, while at later times matter and radiation are negligible. The transition from matter to DE domination is very tight and sharp in $\Lambda$CDM. In the case of PEDE model, as an alternative of $\Lambda$CDM, we can alleviate the coincidence problem. To do this, we perform a comparison between the derivative of $\Omega_{\rm d}$ from Eq.(\ref{eq:omegd}) and that of the standard $\Lambda$CDM model. In Fig.(\ref{fig:figcoin}), we plot the evolution of the derivative of $\Omega_{\rm d}$ with respect to scale factor $a$ ($d\Omega_{d}/da$) as a function of $\log_{10}{a}$. We see that the behavior of $d\Omega_{d}/da$ for PEDE model is completely different from that of the $\Lambda$CDM model. As we know, for both of the models, the energy density of DE at early times is negligible, while at later times it is dominated. In $\Lambda$CDM model, $\Omega_{\rm d}$ changes very slowly all over the time, except at a brief epoch around present time, $a\sim1$. But in PEDE model we observe a completely different behavior. $\Omega_{\rm d}$ in PEDE model changes very faster than energy density of cosmological constant at early times. It can also change in a wide range of scale factor and therefore one can say that PEDE model, at least, alleviates the coincidence problem. 

In order to obtain the equation of state (EoS) parameter for PEDE model, we start with the conservation equation of DE as follows
 \begin{eqnarray}\label{continuity}
&&\dot{\rho_{\rm d}}+3H(1+w_{\rm d})\rho_{\rm d}=0\;\label{eq:de},
 \end{eqnarray}
where over-dot indicates the derivative with respect to cosmic time $t$. Combining Eqs.(\ref{eq:omegd}), (\ref{eq:hub}) \& (\ref{eq:de}), the EoS parameter of PEDE model is obtained as follows:
\begin{eqnarray}\label{eq:wd}
w_{\rm d}(z)=-1+\dfrac{1+z}{3}\dfrac{d\ln \Omega_{\rm d} }{dz}\;.
\end{eqnarray}
By using Eq.(\ref{eq:omegd}), the above equation is written as:
\begin{eqnarray}\label{eq:wd2}
w_{\rm d}(z)=-1-\dfrac{1}{3\ln 10}\left[ 1+\tanh (\log_{10} (1+z))\right]\;.
\end{eqnarray}
From Eq.(\ref{eq:wd2}), we can see that in PEDE model, the EoS of 
DE evolves in phantom regime all over the time. It changes from $w_{\rm d}=-1-\dfrac{2}{3\ln 10}$ at high redshifts to its upper value, $w_{\rm d}=-1$ at $z=-1$ in the far future.

\begin{figure} 
	\centering
	\includegraphics[width=8cm]{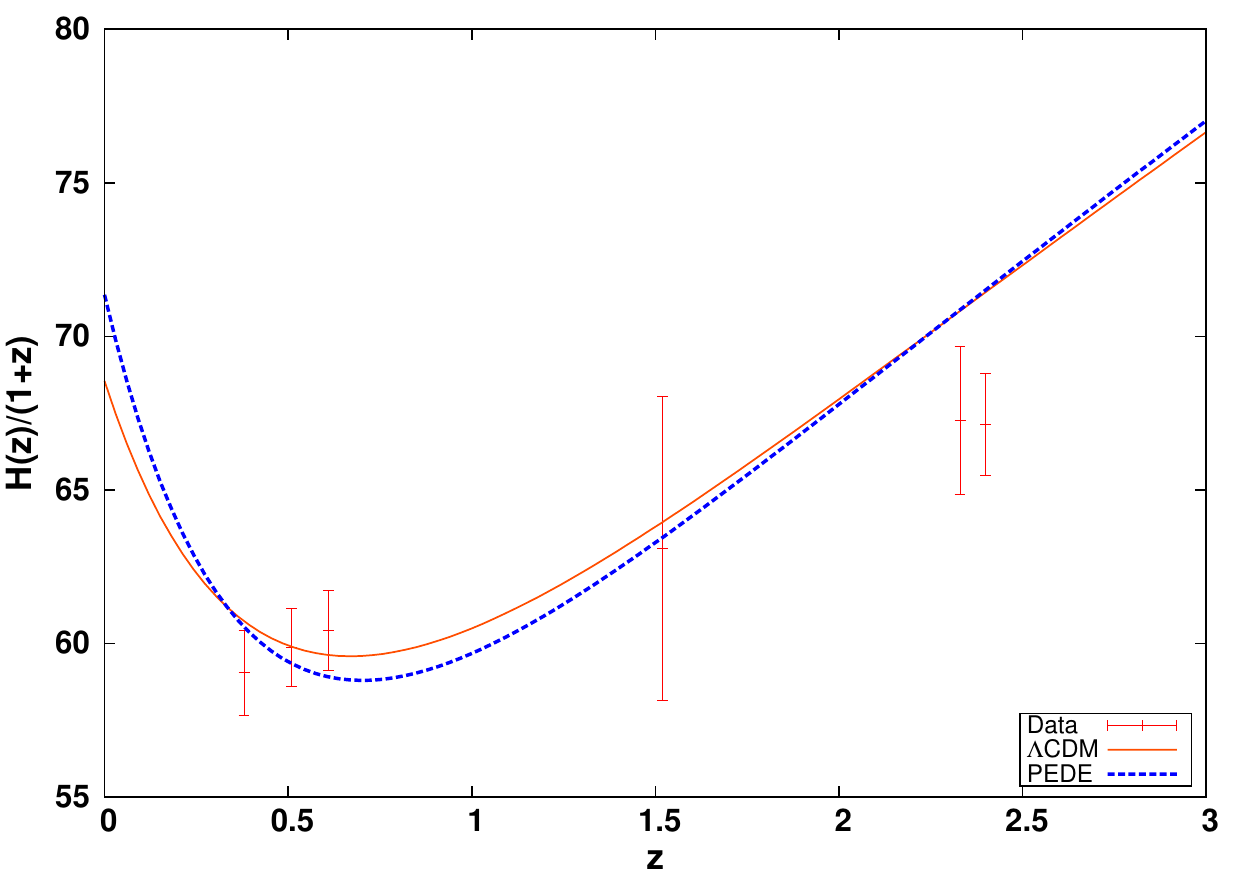}
	\caption{ The redshift evolution of $H(z)/(1+z)$ for PEDE and $\Lambda$CDM and related observational data points. We set $\Omega_{\rm m,0}$ and $H_0$ upon on their best values in last columns of Tables (\ref{tab:pf}) \& (\ref{tab:lf}), respectively, for PEDE and $\Lambda$CDM cosmologies.}
	\label{fig:hz}
\end{figure}
 
\begin{figure} 
	\centering
	\includegraphics[width=8cm]{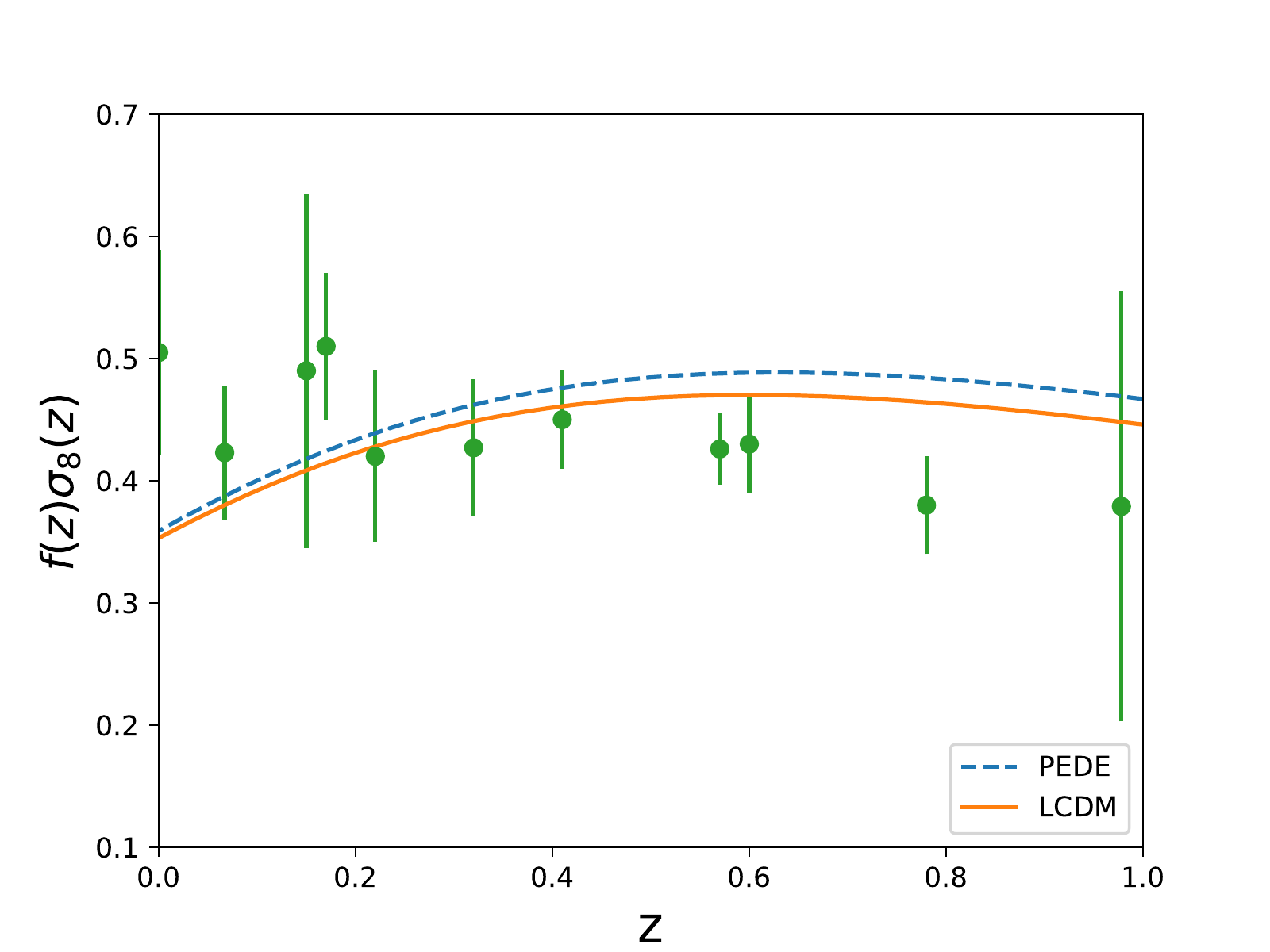}
	\caption{ The redshift evolution of $f\sigma_8$ for PEDE and $\Lambda$CDM and the observational data points. We set the free parameters upon their best fit values from the last columns of Tables (\ref{tab:pf}) \& (\ref{tab:lf}).}
	\label{fig:fs8}
\end{figure}

In order to study the evolution of PEDE model at background level, we have plotted the evolution of $H(z)/(1+z)$ versus redshift in Fig.(\ref{fig:hz}). Here we fix the free parameters $\Omega_{m0}$ and $h$ based on the best fit values in Tables (\ref{tab:pf}) \& (\ref{tab:lf}) for flat PEDE and $\Lambda$CDM models, respectively. We also show some relevant observational data points including BAO measurements from BOSS DR12 \citep{Alam:2016hwk}, data point from BOSS DR14 quasars \citep{Zarrouk:2018vwy} and  BAO measurements from BOSS Ly-$\alpha$ \citep{Bautista:2017zgn,Bourboux:2017cbm}. It is easy to see that the PEDE model and $\Lambda$CDM cosmology have the same behavior in redshift evolution of Hubble parameter.\\ 
In next step, we investigate the PEDE model in perturbation level. DE not only accelerates the expansion of the universe but also changes the growth rate of the large scale structure formation in the universe. Hence, studying DE in perturbation level can help us to distinguish different DE models. For complete review and details, we refer the reader to \citep{Rezaei:2017yyj,Malekjani:2018qcz,Rezaei:2019roe}. Here, in order to improve our knowledge about PEDE model, we investigate the model in perturbation level. In this way, we compute the growth rate function ($f\sigma_8$) as an observable parameter for both of PEDE and $\Lambda$CDM models. The equations for the evolution of growth rate function $f\sigma_8$ can be fond in \citep{Malekjani:2016edh}. 
In Fig.(\ref{fig:fs8}), we plot the redshift evolution of $f(z)\sigma_8$ for both flat PEDE and $\Lambda$CDM models. The data showed in the figure are the latest observational $f\sigma_8$ data reported in next section. Note that the free parameters of models are fixed based on the best fit values reported in Tables (\ref{tab:pf}) \& (\ref{tab:lf}) for flat PEDE and $\Lambda$CDM models, respectively. In overall, one can see that the PEDE model can not fit the observational data in cluster scales as much as the standard $\Lambda$CDM model. Especially, the predicted $f\sigma_8$ for PEDE model deviates from $\Lambda$CDM scenario at higher redshifts.  In next section, using the statistical Bayesian evidence tool, we compare the PEDE and $\Lambda$CDM models with observational data and discuss which of them is in better agreement with observations technically.
 
\section{Observational data and Bayesian inference}\label{sec:be}

In contrast to the maximum likelihood estimator, Bayesian inference not only determines the free parameters but also provides a direct way to compare different models. In this section, we briefly review basic formalism of Bayesian inference and after introducing our observational data, we  compute the Bayesian evidence in different scenarios which we consider in this paper.
 
The Bayes theorem is given by a simple relation:
\begin{eqnarray}\label{be1}
p(A\mid B)=\dfrac{p(B\mid A) p(A)}{p(B)}\;,
\end{eqnarray}

considering A as the free parameters $(\Theta)$ and B as data set $(\cal D)$, we have
\begin{eqnarray}\label{be3}
p(\Theta\mid {\cal D}, {\cal M})=\dfrac{ p({\cal D}\mid \Theta , {\cal M}) p(\Theta \mid {\cal M})}{p({\cal D} \mid {\cal M})}\;,
\end{eqnarray}
where the model $\cal M$ has been shown explicitly. This relation simply tells us having the likelihood $(p({\cal D}\mid \Theta ))$ and prior $(p(\Theta))$, we can compute the posterior distribution on $\Theta$ $(p(\Theta\mid {\cal D}))$. The denominator in Eq.(\ref{be3}) is given by: 
\begin{eqnarray}\label{be4}
\varepsilon = p({\cal D}\mid {\cal M})=\int p(\Theta \mid {\cal M}) p({\cal D} \mid \Theta , {\cal M}) d\Theta\;,
\end{eqnarray}
and called the Bayesian evidence or marginal likelihood. Although this might has an analytic solution for a low dimensional cases, for a high denominational problem it is intractable analytically and one has to use numerical methods to evaluate the integral. In this paper, we use the Sequential Monte Carlo (SMC) algorithm to sample the posterior. Notice that the Bayesian evidence is a by-product of the SMC method.

The evidence is a crucial quantity for model selection in Bayesian framework and in comparison between two models. The model with higher evidence is favored over another one. Moreover, the Bayesian evidence for model selection has been widely used in cosmology  \citep{Hobson:2002zf,Saini:2003wq,Parkinson:2006ku,Martin:2010hh,Lonappan:2017lzt,Mehrabi:2018oke}.   In this paper, we use the Jeffreys' scale \citep{jeff.book} to measure the significant difference between two models.  To do this, having two models $M_1$ and $M_2$  the Jeffreys scale with respect of $\Delta\ln \varepsilon = \ln \varepsilon _{M_1}- \ln \varepsilon _{M_2}$ is as the following: \citep{Nesseris:2012cq}:
\begin{itemize}
	\item for $\Delta\ln \varepsilon < 1.1$ there is a weak evidence against model $M_2$.
	\item for $1.1<\Delta\ln \varepsilon < 3$ there is a definite evidence against model $M_2$.
	\item and finally for $3<\Delta\ln \varepsilon$ there is a strong evidence against model $M_2$. 
\end{itemize}

In addition to the evidence, there are also other measurements to compare models. Among these quantities, we compute the Akaike Information (AIC)\citep{Akaike:1974} and Bayesian Information Criterion (BIC)\citep{Schwarz:1974}. These measurements are given by:
\begin{eqnarray}\label{aic}
AIC=\chi_{min}^2+2k\;,\\
BIC=\chi_{min}^2+k \ln N\;.
\end{eqnarray}
where $k$  ($N$) is the number of fitting parameters( number of data points).

\begin{table}
 \centering
 \caption{The ranges of the model parameters which we consider in this work as the prior. We note that we assumed uniform priors for all of the model parameters. Notice that $\Omega_{\rm dm0}$ and $\Omega_{\rm bm0}$ represent the present values of energy densities for dark matter and baryons, respectively. The energy density of total non-relativistic matter is sum of dark matter and baryons as $\Omega_{\rm m}=\Omega_{\rm dm}+\Omega_{\rm bm}$. 
}
\begin{tabular} {c c c c}
\hline
 Parameter & Prior & \\
\hline
{\boldmath$\Omega_{\rm dm0}       $} & $0.15 - 0.35$  \\

{\boldmath$\Omega_{\rm bm0}       $} & $0.03 - 0.06$  \\

{\boldmath$\Omega_{\rm d0}       $}  & $0.05 - 1.20$ &(In the case of non-flat universe) \\

{\boldmath$h              $} & $0.6 - 0.8$  \\

{\boldmath$\sigma_8       $} & $0.6 - 1.2$\\
\hline
\end{tabular}\label{tab:prior}
\end{table}
Since to compute the Bayesian evidence the prior is relevant, we show the prior for each of model parameter in Tab.(\ref{tab:prior}) for both flat and non-flat universes. Notice that $\Omega_{\rm m0}$ and $\Omega_{\rm d0}$ in Tab.\ref{tab:prior} represent the current values of non-relativistic matter and DE, respectively. In a general non-flat universe $\Omega_{\rm d}=1-\Omega_{\rm m}-\Omega_{K}$ and in a flat universe, it reduces to $\Omega_{\rm d}=1-\Omega_{\rm m}$. 

Before going through the details of our analysis, in the following, we first introduce our observational data set. We use the following background and perturbation cosmological data.    
\begin{itemize}
	\item CMB distance prior from final Planck 2018 release \citep{Chen:2018dbv}
	\item Latest measurements from cosmic chronometers for $H(z)$ from \citep{Farooq:2016zwm} 
	\item The SnIa sample from the Pantheon sample \citep{Scolnic:2017caz}
	\item  5 measurements of $D_V(z)$  from WiggleZ \citep{Blake:2011en} (three data) ,6df Galaxy \citep{Beutler:2011hx} and MGS \citep{Ross:2014qpa}
	\item BAO signal measurements with their full covariance matrix from BOSS DR12 \citep{Alam:2016hwk}
	\item Radial and transverse BAO measurements from Ly$\alpha$-Forests SDSS DR12 \citep{Bautista:2017zgn}
	\item radial and transverse BAO measurements  from quasar sample from BOSS DR14 \citep{Gil-Marin:2018cgo}
	\item Measurement of the angular diameter distance from DES Collaboration \citep{Abbott:2017wcz}
	\item Local Measurement of the Hubble constant $H_0$ \citep{Riess:2019cxk}
	\item Measurements on the baryons from Big Bang nucleosynthesis (BBN) as $100\Omega_{\rm bm0}h^2=2.235$. \citep{Cooke:2017cwo}.
	\item The $f\sigma_8$ data extracted from RSD data including: a data point from 2dFGRS \citep{Song:2008qt}, four data points from WiggleZ \citep{Blake:2011rj}, one data point from 6dFGRS \citep{Beutler:2012px}, one data point from SDSS Main galaxy sample \citep{Howlett:2014opa}, one data point from 2MTF \citep{Howlett:2017asq}, two data points from BOSS DR12 \citep{Gil-Marin:2016wya}, one data point from FastSound \citep{Okumura:2015lvp} and finally two data points from eBOSS DR14 \citep{Zhao:2018jxv}.
\end{itemize}

We note that some data points of the $H(z)$ measurements reported in \citep{Farooq:2016zwm} are obtained from the same BAO observations. Notice that, because of their overlap to BAO data points, we can not use them beside BAO data. Thus we remove these $H(z)$ data points from our data samples.
Having the mentioned statistical tools, we consider three different steps. First we use all of background datasets except CMB data and find the posterior distribution of parameters through the SMC algorithm. Then, we add the CMB data to investigate the effect of high redshift data and finally we use all background data jointed to the growth rate data. Concerning the growth rate data, we should note that the cosmic surveys do not measure distances to galaxies directly. Hence, one should assume a specific cosmological model. The observational growth rate datapoints are reported in the context of flat $\Lambda$CDM cosmology. To resolve this model dependence, we should use a correction before applying these data points in our analysis \citep[see][for more details]{Macaulay:2013swa,Alam:2015rsa,Nesseris:2017vor,Shafieloo:2018gin}. The correction factor can be obtained by calculating  the ratio of $H(z)D_A(z)$ of the cosmology used to that of the standard flat $\Lambda$CDM cosmology \citep{Macaulay:2013swa}. Although this correction itself is quite small, we implement it as follows. First, we obtain the correction factor $(CF)$ as the ratio of the product of the $H(z)$ and the angular diameter distance $d_A(z)$ for the model at hand to that of the fiducial cosmology:

\begin{eqnarray}\label{eq:corfactor}
CF(z)= \dfrac{H(z)d_A(z)}{H^{fid}(z)d^{fid}_A(z)}\;.
\end{eqnarray}
where the values of the fiducial cosmology can be found in data point references. Now, using correction factor $CF$ and multiplying it on the theoretical prediction of $f(z)\sigma_8(z)$, we can be sure about the independency of our datapoints from the cosmological models.
We run our code several times with different initial sample points to check the stability of both our results and evidences in each case.  In the next section, we will present the results of our analysis.

\section{Results and discussion}\label{sec:rd}
In the case of flat- PEDE model, we show the best values of free parameters alongside their $1\sigma$ uncertainties in Tab.(\ref{tab:pf}). Here, we have used three different combinations of observational data sets. In the first column, we combine SnIa, BAO, H(z), BBN, and local $H_0$ data points (background data without CMB), while in the second column we add CMB data to the previous ones (background data with CMB). Finally in the third column, we joint the growth rate data to all background data. In the same way, we repeat our analysis for flat- $\Lambda$CDM model and report the results in Tab.(\ref{tab:lf}).

\begin{table}
 \centering
 \caption{The best fit value of free parameters and $1\sigma$ uncertainties
using different data sets for PEDE model in flat universe.
}
\begin{tabular} { l  c c c c}
\hline
 Parameter & & 68\% limits &\\
\hline
  & without CMB & with CMB & with $f\sigma_8$\\
\hline
{\boldmath$\Omega_{\rm m0}       $} & $0.2909\pm 0.0086 $& $0.2860\pm 0.0050$  &  $0.2872\pm 0.0051$\\

{\boldmath$h              $} & $0.7087\pm 0.0065 $& $0.7137\pm 0.0045$  & $0.7126\pm 0.0045$\\

{\boldmath$\sigma_8       $} & -- & --  & $0.867\pm 0.030$\\
\hline
\end{tabular}\label{tab:pf}
\end{table}

\begin{table}
 \centering
 \caption{The best fit value of free parameters and $1\sigma$ uncertainties
using different data sets for $\Lambda$CDM model in flat universe.
}
\begin{tabular} { l  c c c c }
\hline
 Parameter & & 68\% limits &\\
\hline
  & without CMB & with CMB & with  $f\sigma_8$\\
\hline
{\boldmath$\Omega_{\rm m0}       $} & $0.2883\pm 0.0088$& $0.3019\pm 0.0051$  & $0.3025\pm 0.0052$\\

{\boldmath$h              $} & $0.6875\pm 0.0063$& $0.6851\pm 0.0040$  & $0.6846\pm 0.0040$\\

{\boldmath$\sigma_8       $} & -- & --  & $0.823\pm 0.028$\\
\hline
\end{tabular}\label{tab:lf}
\end{table}

We can see from Tabs. (\ref{tab:pf} \& \ref{tab:lf}), different combinations of datasets yield different values of best fit parameters. However, the differences are up to about $1\sigma$ uncertainty of the best fit parameters. From the second rows of tables, we observe that the best fit value of $H_0=100h$ reported for PEDE model is higher than that of the concordance model in the light of alleviating the tension between the Planck inferred value $H_0=67.4\pm0.5$ km/s/Mpc \citep{Aghanim:2018eyx} and the local measurement
value  $H_0=74.03\pm1.42$ km/s/Mpc by Riess et al. \citep{Riess:2019cxk}. Quantitatively speaking, our results for flat- PEDE universe show roughly $2-3\sigma$ decrement of tension for different combinations of datasets in Tab. (\ref{tab:pf}). Notice that the tension between flat- $\Lambda$CDM model and the local measurement from Riess et al. \citep{Riess:2019cxk} is approximately $4\sigma$ for all combinations of datasets in Tab. (\ref{tab:lf}).  

\begin{table}
 \centering
 \caption{The best fit values of free parameters and $1\sigma$ uncertainties
using different data sets for PEDE model in non-flat universe.
}
\begin{tabular} { l c c c c}
\hline
 Parameter & & 68\% limits &\\
\hline
  & without CMB & with CMB & with  $f\sigma_8$\\
\hline
{\boldmath$\Omega_{\rm m0}       $} & $0.2864\pm 0.0097$ & $0.2874\pm 0.0052$ &$0.2890\pm 0.0051$\\

{\boldmath$\Omega_{\rm d0}       $} & $0.680\pm 0.028$ & $0.7149\pm 0.0050$ &$0.7133\pm 0.0050$\\

{\boldmath$h              $} & $0.700\pm 0.010$ & $0.7082\pm 0.0057$ & $0.7069\pm 0.0056$\\

{\boldmath$\sigma_8       $} & -- & -- & $0.865\pm 0.030$\\
\hline
\end{tabular}\label{tab:pnf}
\end{table}

\begin{table}
 \centering
 \caption{The best fit values of free parameters and $1\sigma$ uncertainties
using different data sets for $\Lambda$CDM model in non-flat universe.
}
\begin{tabular} { l c c c c}
\hline
 Parameter & & 68\% limits &\\
\hline
  & without CMB & with CMB & with  $f\sigma_8$\\
\hline
{\boldmath$\Omega_{\rm m0}       $} & $0.2981\pm 0.0098$ & $0.2997\pm 0.0053$ &$0.3004\pm 0.0052$\\

{\boldmath$\Omega_{\rm d0}       $} & $0.788\pm 0.032$ & $0.6978\pm 0.0052$ &$0.6972\pm 0.0051$\\

{\boldmath$h              $} & $0.707\pm 0.010$ & $0.6910\pm 0.0054$ & $0.6902\pm 0.0054$\\

{\boldmath$\sigma_8       $} & -- & -- & $0.825\pm 0.028$\\
\hline
\end{tabular}\label{tab:lnf}
\end{table}

Now we present the results of our analysis for non-flat PEDE and  $\Lambda$CDM model models, respectively, in Tabs. (\ref{tab:pnf} \& \ref{tab:lnf}).
Same as the flat geometry, in the case of non-flat PEDE model, we obtain the larger value of $H_0$ compared to non-flat $\Lambda$CDM model. Hence the tension of $H_0$ for non-flat PEDE model is lower than the non-flat $\Lambda$CDM universe. Notice that our results for $\sigma_8$ quantity indicate that the present value of $\sigma_8$ for both flat and non-flat PEDE model is roughly \textbf{$1.1\sigma$} larger than that of the $\Lambda$CDM model. Hence comparing the results of low-redshift observations with Planck inferred value of $\sigma_8$, one can say the concordance $\Lambda$CDM model is in better situation than the PEDE models (for more details, see Tables(\ref{tab:pf}-\ref{tab:lnf})).\\

Concerning the curvature parameter of the universe, $\Omega_{k0}$, for the models under study we get \textbf{$\Omega_{k0} = 0.0024^{+0.0015}_{-0.0015}$} for non-flat $\Lambda$CDM obtained from the analysis based on all datasets. For non-flat PEDE model we have \textbf{$\Omega_{k0} = -0.0023^{+0.0014}_{-0.0014}$}. While a spatially flat universe is strongly supported by different cosmological probes, we can see that the non-flat PEDE model meets $\Omega_{k0}=0$ at \textbf{$1.6\sigma$}. In the case of non-flat  $\Lambda$CDM, we obtain approximately the same value of tension with flat universe. In this case we have $\Omega_{k0}=0$ at $\sim 1.6\sigma$. 

\begin{table*}
	\centering
	\caption{The statistical results of the analysis for different cosmologies considered in this work using different data sets.
	}
	\begin{tabular} { l  c c c c c c c  c  c c}
		\hline
		Data & Curvature & Model & $\chi_{min}^2$ & AIC & $\Delta AIC$ & BIC & $\Delta BIC$ & $\ln \varepsilon$ &  $ \Delta \ln \varepsilon = \ln \varepsilon_{flat-\Lambda CDM} - \ln \varepsilon_{Model} $\\
		\hline
		&  & $\Lambda$CDM         & 94.02 & 98.02 & 0.0& 103.0  & 0.0 & -52.48 & 0.0 \\
		&  Flat    & PEDE & 100.48 & 104.48 & 6.46 & 109.46  & 6.46 & -54.40 & 1.92 \\
		Without CMB& & & & & &  & & \\
		&  Non-Flat    & $\Lambda$CDM & 92.40 & 98.40 & 0.38 & 105.86 & 2.87 & -54.21 & 1.73 \\
		& & PEDE & 98.44  & 104.44 & 6.42 & 111.90 & 8.91 & -56.83 & 4.35 \\
		\hline
		& & $\Lambda$CDM & 100.86 & 104.86 & 0.0 & 109.90 & 0.0 & -59.31 & 0.0 \\
		&   Flat    & PEDE & 102.55 & 106.55 & 1.69 & 111.59 & 1.69 & -59.63 & 0.32  \\
		With CMB& & & & & &  & &  \\
		&  Non-Flat   & $\Lambda$CDM & 99.55 & 105.55 & 0.69 & 113.11 & 3.21 & -65.27 & 5.96 \\
		&  & PEDE & 100.93 & 106.93 & 2.07 & 114.49 & 4.59 & -66.27 & 6.97  \\
		\hline
		&  & $\Lambda$CDM &111.84 & 117.84 & 0.0 & 125.86 & 0.0 & -65.58 & 0.0 \\
		&   Flat   & PEDE & 123.85 & 129.85 & 12.01 & 137.87 & 12.01 & -71.25 & 5.66 \\
		With $f\sigma_8$& & & & & &  & & \\
		& Non-Flat    & $\Lambda$CDM & 110.28 & 118.28 & 0.44 & 128.97 & 3.11 & -71.19 & 5.61  \\
		&  & PEDE & 122.33 & 130.33 & 12.49 & 141.02 & 15.16 & -77.17 &  11.58\\
		
		\hline
	\end{tabular}\label{tab:result}
\end{table*}
Finally, we compare different models in each geometry studied in our analysis. Notice that both of PEDE and $\Lambda$CDM models have the same number of free parameters in the same geometry and therefore for comparisons between models assuming the $\chi_{min}^2$ values is sufficient. However, when we compare the flat and non-flat cosmologies, because of different numbers of free parameters, we should compute the information criteria( AIC and BIC) beside the Baysian evidence parameter. The results of our analysis are showed in Tab.(\ref{tab:result}).
Using the value of AIC criteria, we can select the best model-data fit as follows. For the combination of background datasets without CMB, the flat- $\Lambda$CDM has the minimum value of AIC number. In this case we can find consistency between flat and non-flat $\Lambda$CDM models because of $\Delta AIC < 2$. While for PEDE model in both of flat and non-flat universes, $\Delta AIC > 4$ indicating a positive evidence against these models. Adding the CMB data to previous background datasets, leads to better results for PEDE model. In this step, the value of $\Delta AIC$ for both flat- PEDE and non-flat $\Lambda$CDM models are smaller than 2. Thus we can say that the flat- PEDE and concordance flat- $\Lambda$ CDM and non-flat $\Lambda$CDM scenarios are consistent with each other. On the other hand, we have $\Delta AIC >2$ for non-flat PEDE model, representing no significant support for this model. Finally in the case of combinations of all background and growth data, we get essentially no support ($\Delta AIC > 10$) for  both of flat and non-flat PEDE models, while we have significant support to the non-flat $\Lambda$CDM because in this case we obtain $\Delta AIC=0.44$. 

Using the value of BIC criteria, we present our results as follows. In the case of the combinations of background datasets without CMB, we conclude that the flat- $\Lambda$CDM model is the best model. There is a positive evidence against non-flat $\Lambda$CDM models and strong evidence ($6<\Delta BIC <10$) against both flat- PEDE non-flat PEDE cosmology. In the case of combined background datasets with CMB data, the results get better for PEDE cosmology. We see that in the case of flat- PEDE model $\Delta BIC <2$ meaning that there is a weak evidence against this model. We also observe a positive evidence against both the $\Lambda$CDM and PEDE models in non-flat geometry. Eventually in the case of combined background datasets with growth rate data, there is a positive and very strong evidence against non-flat $\Lambda$CDM as well as both flat and non-flat PEDE scenarios.  

Finally, we report the result of Bayesian evidence analysis. In the case of background datasets without CMB, we have $1.1<\Delta \ln \varepsilon < 3.0$ meaning the definite evidence against flat- PEDE and non-flat $\Lambda$CDM models. In the case of non-flat PEDE model, we get $\Delta \ln \varepsilon > 3.0$ meaning the strong evidence against the model. In the case of combined background datasets with CMB data, we conclude that the flat- PEDE model is well consistent with flat $\Lambda$CDM model ($\Delta \ln \varepsilon < 1.1$). While we obtain the strong evidence against both of the non-flat PEDE and $\Lambda$CDM cosmologies. Finally in the case of combined all background data with growth rate one, we observe the strong evidence against both flat and non-flat PEDE and also non-flat $\Lambda$CDM models.

\section{Conclusion}\label{sec:c}
In this work, we studied the recent phenomenological emergent dark energy (PEDE) model by \citep{Li:2019yem} using the latest observational data based on the Bayesian inference analysis. The datasets that we used in this work are included from background datasets and growth rate of perturbations from RSD observations. The background datasets used here are: the CMB distance prior from final Planck 2018 \citep{Chen:2018dbv}; SnIa data from Pantheon sample \citep{Scolnic:2017caz}; some of the latest measurements from cosmic chronometers for $H(z)$ \citep{Farooq:2016zwm}; measurements of $D_V(z)$ from WiggleZ \citep{Blake:2011en}, 6df Galaxy \citep{Beutler:2011hx}, MGS \citep{Ross:2014qpa}; BAO measurements from BOSS DR12 \citep{Alam:2016hwk}; Radial and transverse BAO measurements from Ly$\alpha$-Forests SDSS DR12 \citep{Bautista:2017zgn}; Radial and transverse BAO measurements from quasar sample of BOSS DR14 \citep{Gil-Marin:2018cgo}; measurements of the angular diameter distance from DES collaboration \citep{Abbott:2017wcz}; baryon density measurements  derived from BBN \citep{Cooke:2017cwo}; local measurements on the Hubble constant $H_0$ \cite{Riess:2019cxk}. The growth rate data used in our analysis are the $f\sigma_8$ data extracted from RSD data from 2dFGRS \citep{Song:2008qt}, WiggleZ \citep{Blake:2011rj}, 6dFGRS \citep{Beutler:2012px}, SDSS Main galaxy sample \citep{Howlett:2014opa}, 2MTF \citep{Howlett:2017asq}, BOSS DR12 \citep{Gil-Marin:2016wya}, Fast Sound \citep{Okumura:2015lvp} and finally eBOSS DR14 observations \citep{Zhao:2018jxv}.

One of the important note in the modern cosmology is determining the spatial curvature of the universe using observations. As a part of our analysis, we  examined the presence of cosmic curvature and studied the effect of curvature parameter on the fitting data. To do this,  we assumed the non-flat universe (by adding a new free parameter) for both of the models under study. In addition to Bayesian evidence, we applied the relevant AIC and BIC criteria to compare the flat and non-flat cosmologies. Although a spatially flat cosmology is strongly favored by different cosmological observations, but some of studies have shown that fitting cosmological observations to dynamical DE models can satisfy a non-flat universe \citep{Li:2016wjm,Wei:2016xti}. So in this work we analyzed the following cosmological models:
(i). The flat - $\Lambda$CDM model.
(ii). Non-flat $\Lambda$CDM model.
(iii). The flat - PEDE model.
(iv). Non-flat PEDE model.\\
We assumed three different combinations of datasets in our analysis. Firstly, we used all of background datasets except CMB one. Secondly, we added the CMB data to investigate its effect on our results and thirdly we used all background data combined with the growth rate observations. Our main conclusions in this work are as follows.\\
\begin{itemize}
	\item  {\bf Tensions of $H_0-\sigma_8$}: our results showed that the PEDE models can decrease $2 \sim 3 \sigma$ of the tension of $H_0$ appearing in concordance $\Lambda$CDM model. However, the PEDE models can not alleviate the tension of $\sigma_8$ between low-redshift observations and Planck inferred value. Concerning the $H_0$ tension, our results obtained based on the different combinations of datasets are in agreement with those of \citep{Pan:2019hac}. In fact one of the interesting properties of the PEDE model is that it can alleviate the  $H_0$ tension, without adding any degree of freedom.
	\item {\bf Non-flatness of the universe}: The constraint results showed that a positive $\Omega_{k0}$ is preferred by non-flat $\Lambda$CDM model. In this case,  because of the large variance, the flat $\Lambda$CDM is not ruled out to more than \textbf{$1.6 \sigma$} region. For the non-flat PEDE model, our analysis leads to a negative value for $\Omega_{k0}$ which can meet $\Omega_{k0}=0$ at \textbf{$\sim 1.6 \sigma$} confidence level.
	\item  {\bf Model selection:} (i). For the combination of background datasets without CMB, all AIC, BIC and Bayesian evidence analysis showed that the flat-$\Lambda$CDM model is the best model, positive evidence against flat- PEDE and non-flat $\Lambda$CDM model and eventually strong evidence against non-flat PEDE model. (ii). For the combination of all background datasets (including CMB), all of the three analysis show that the flat- PEDE model is well consistent with observations as much as the best model. While we observed the positive and strong evidence against both of non-flat PEDE and standard models. This result is in agreement with the results of Planck 2018 \citep{Aghanim:2018eyx} which indicates that the spatial curvature of our universe is consistent with a flat geometry. (iii). Finally for the combinations of all background data with the growth rate dataset, our analysis showed that there is a positive evidence against the non-flat $\Lambda$CDM scenario and very strong evidence against both of the flat and non-flat PEDE cosmologies. So we can conclude that the PEDE models cannot fit the observations in cluster scales as equally as standard $\Lambda$CDM model.   
\end{itemize}

\section{Acknowledgements}
The works of M. Rezaei and T. Naderi have been supported financially by Iran Science Elites Federation. The authors should thank the anonymous reviewer for him/her careful reading of the manuscript and giving the valuable comments and suggestions.

  \bibliographystyle{apsrev4-1}
\bibliography{ref}

\end{document}